\documentstyle[12pt]{article}
\begin{document}
\begin{flushright}
TIFR/TH/96-59\\
hep-ph/9610505\\
\end{flushright}
\vskip24pt
\begin{center}
{\LARGE\bf QUARKONIUM PRODUCTION VIA FRAGMENTATION: A REVIEW 
\footnote{Presented at the Fourth Workshop on High Energy Physics 
Phenomenology (WHEPP IV), Calcutta, India, January, 1996.}}\\
\end{center}
\vspace{10mm}
\noindent
{\Large K. Sridhar, Theory Group, Tata Institute of Fundamental Research,
Homi Bhabha Road, Bombay 400 005, India} \\
\vspace{10mm}

\begin{abstract}
We review the physics of quarkonium production at high 
energy colliders, with a specific emphasis on production 
$via$ fragmentation. We discuss the fragmentation picture
for the production of quarkonia at large-$p_T$ at the 
Tevatron and emphasise the importance of the colour-octet
components of the quarkonium wave function which are included
in the fragmentation functions. Applications of these theoretical
ideas to other processes, both at the Tevatron and at other
experiments like LEP, HERA and the LHC are important as they
will provide stringent tests of these ideas.
\end{abstract}
\vskip 40pt
The production of quarkonia has conventionally been described in 
terms of the colour-singlet model \cite{berjon, br}. In this model,
a non-relativistic approximation is used to describe the binding
of the heavy-quark pair produced via parton-fusion processes, into
a quarkonium state. The heavy-quark pair is projected onto a 
physical quarkonium state using a colour-singlet projection and 
an appropriate spin-projection. It is also necessary to account 
for $J/\psi$'s produced $via$ electromagnetic decays of the 
$P$-states, $viz.$, the $\chi$'s, since inclusive $J/\psi$ production 
is measured in experiments. In the model, both the direct 
$S$-state production and the indirect production $via$ $P$-state 
decays are accounted for. This model has been successfully applied \cite{br} 
to describe large-$p_T$ $J/\psi$ production at the relatively low 
energies in the ISR experiment. In going from the ISR energy 
($\sqrt{s}=63$~GeV) to the UA1 energy ($\sqrt{s}=630$~GeV), it was
found \cite{gms} that the production of $b$-quarks and their subsequent
decay becomes an important source of $J/\psi$
production. While at the UA1 experiment it was not possible to
separate the $b$-quark contribution, the use of the silicon-vertex
detector at the CDF experiment in studying $J/\psi$ production
in $\bar p p$ collisions at the Tevatron ($\sqrt{s}=1.8$~TeV)
allows the subtraction of the $b$-quark contribution from the
total yield, thereby providing a measurement of direct
inclusive $J/\psi$ production at large-$p_T$.
The inclusive $J/\psi$ production cross-section measured by the
CDF experiment \cite{cdf} turned out to be an order of magnitude
larger than the prediction of the colour-singlet model. 

In fact, even before the experimental results from the CDF 
collaboration were available, it was shown
by Braaten and Yuan \cite{bryu} that in addition to the parton
fusion contributions taken into account in the colour-singlet
model, fragmentation of gluons and charm quarks could be an
important source of large-$p_T$ $J/\psi$ production. In the
fragmentation process, one considers the production of charmonia 
from final-state gluons or charm quarks which have large $p_T$, 
but almost zero virtuality. The fragmentation contribution is 
computed by factorising the cross-section for the process 
$AB \rightarrow (J/\psi,\chi_J) X$ (where $A,\ B$ denote hadrons)
into a part containing the hard-scattering cross-section for producing a
gluon or a charm quark and a part which specifies the fragmentation of
the gluon or the charm quark into the required charmonium state, i.e.
\begin{equation}
d\sigma (AB \rightarrow (J/\psi,\chi_J) X)
 = \sum_i \int_0^1 dz \hskip4pt
d\sigma (AB \rightarrow i X) D_{i \rightarrow (J/\psi,\chi_J)}
(z,\mu ) ,
\label{e1}
\end{equation}
where $i$ is the fragmenting parton (either a gluon or a charm quark).
$D(z,\mu)$ is the fragmentation function and $z$ is
the fraction of the momentum of the parent parton carried by the
charmonium state. Because the gluon or the $c$-quark fragments into
a heavy quarkonium state, the fragmentation function can be computed
perturbatively, in the same spirit as in the colour-singlet model.
This yields the fragmentation function at an initial scale $\mu_0$
which is of the order of $m_c$. If the scale $\mu$ is chosen to be
of the order of $p_T$, then large logarithms in $\mu/m_c$ appear
which have to be resummed using the usual Altarelli-Parisi equation:
\begin{equation}
\mu {\partial \over \partial\mu} D_{i\rightarrow (J/\psi,\chi_i)} 
(z) = \sum_j\int_{z}^1{dy \over y} P_{ij}({z\over y},\mu)
D_{j\rightarrow (J/\psi,\chi_i)}(y) ,
\label{e2}
\end{equation}
where the $P_{ij}$ are the splitting functions of a parton $j$
into a parton $i$. Only the fragmentation of gluons and
charm quarks need be considered since the light quark contributions 
are expected to be very small. 

The fragmentation function at the initial scale for the production of 
a quarkonium state ${\cal O}$ from a gluon, $D_{g \rightarrow {\cal O}} 
(z,2m_c)$, can be obtained by factorising the process $gg \rightarrow
{\cal O} + n \ {\rm gluons}$ into two parts. The first part specifies 
the production of a large-$p_T$ but low-virtuality gluon. In the limit
that the virtuality can be completely neglected with respect to the
$p_T$ of the gluon, this factor reduces to the on-shell scattering
amplitude for the process $gg \rightarrow gg$. The second part
is the amplitude for the process $g^* \rightarrow {\cal O} + n-1 \ {\rm 
gluons}$. The real gluon production process has the form ${\cal M}_
{\alpha}\epsilon^{\alpha}(q)$, whereas the fragmentation process has
the form ${\cal M}_{\alpha} (-ig^{\alpha\beta}/q^2){\cal A}_{\beta}$,
where the $q^2$ in the denominator is of the order of $m_c^2$.
The fragmentation probability is the ratio of the rates for these
two processes. The fragmentation function is obtained after integrating
over the invariant mass of the fragmenting gluon from $4m_c^2$ to 
infinity. The charm-quark fragmentation functions are also 
calculated in a similar manner. While this calculation goes through
in a straightforward manner for $S$-state quarkonia, for the $P$-states
it is more complicated. This has to with a non-factorising logarithmic
infrared divergence that is encountered in the colour-singlet treatment 
of $P$-state fragmentation functions \cite{bryu2, ma}. The divergence 
arises from soft-gluon radiation that produces the ${}^3P_J$ bound state. 
The problem arises because of assuming that the $c \bar c$ pair can exist
only in a colour-singlet state. This assumption is tantamount to
neglecting the relative velocity $v$ between the $c$ and the $\bar c$.
However $v$ is, in general, not negligible and O$(v^2)$ corrections 
need to be taken into account. A formulation based on non-relativistic 
QCD (NRQCD), using the factorisation method has been recently carried out 
\cite{bbl}, and in this formulation the quarkonium wave-function admits 
of a systematic expansion in powers of $v$ in terms of Fock-space 
components~: for example, the $\chi$ states have the 
colour-singlet $P$-state component at leading order, but there exist 
additional contributions at non-leading order in $v$, which involve octet 
$S$-state components; i.e. 
\begin{equation}
  \vert \chi_J \rangle = O(1) \vert Q\bar Q \lbrack {}^3P_J^{(1)} \rbrack
      \rangle + O(v) \vert Q\bar Q \lbrack {}^3S_1^{(8)} \rbrack g
      \rangle + \ldots
      \label{e3}
\end{equation}
The importance of the colour-octet components had already been seen in
the analysis of the decays of the $\chi$ states \cite{bbl2} where the 
colour-singlet analysis \cite{bgr} revealed a similar problem with the 
logarithmic infrared singularity. But the octet component allows the 
infrared singularity to be absorbed $via$ a wave-function renormalisation, 
without having to introduce an arbitrary infrared cut-off. Thus, a 
consistent perturbation theory for the $P$-states necessiates the inclusion 
of the octet component of the wave-function. It should be pointed out
that the colour-singlet term for the production of a $P$-state
already starts at order $v$, and the colour-octet production $via$
an octet ${}^3S_1$ state also appears at the same order in $v$.

The gluon- and charm-fragmentation functions 
have been calculated \cite{bryu, bryu2, ma, brcyu, chen, yuan}
and using these as inputs several authors \cite{jpsi} have
computed the contributions coming from fragmentation to the total
$J/\psi$ yield and have found that the gluon fragmentation contribution
significantly increases the cross-section, and the order-of-magnitude
discrepancy between the theory and the data from
the CDF experiment can be resolved. The analyses of the CDF data,
thus, demonstrates the importance of the fragmentation mechanism
as an important source of $J/\psi$ production at large $p_T$.

For $S$-state resonances like the $J/\psi$ and the 
$\psi^{\prime}$, the octet contribution is suppressed by powers of
$v$. Further, the $S$-wave amplitude is not infrared divergent 
and can, therefore, be described in terms of a single colour-singlet
matrix-element. But the CDF measurement \cite{cdf2} of
the ratio of $J/\psi$'s coming from $\chi$ decays to those produced
directly shows that the direct $S$-state production is much
larger than the theoretical estimate. It has been suggested \cite{cgmp}
that a colour octet component in the $S$-wave production coming from
gluon fragmentation as originally proposed in Ref.~\cite{brfl}, can
explain this $J/\psi$ anomaly. This corresponds to a virtual gluon 
fragmenting into an octet ${}^3S_1$ state which then makes a double 
E1 transition into a singlet ${}^3S_1$ state. While this process is 
suppressed by a factor of $v^4$ as compared to the colour-singlet process, 
it is enhanced by a factor of $\alpha_s^2$. One can fix the value of the 
colour-octet matrix-element by normalising to the data on direct $J/\psi$ 
production cross-section from the CDF experiment. The colour-octet contribution to $S$-state production has also been invoked \cite{brfl} to explain the 
large $\psi^{\prime}$ cross-section measured by CDF \cite{cdf}, but there 
can be a sizeable contribution to this cross-section coming from the
decays of radially excited $P$-states \cite{psip}.

The long-distance octet matrix elements are not calculable, and 
have to be fitted by comparing with various experiments. After
the suggestion of Ref.~\cite{brfl} that the $S$-state production
could also be dominated by octet components, the production of
charmonium and bottomonium resonances as direct octet production
amplitudes via parton fusion has been computed in Ref.~\cite{cho}, 
rather than using the fragmentation approximation which holds only 
for large-$p_T$. It is found that for $p_T \ge 10$~GeV the 
fusion production of quarkonia $via$ octet channels matches quite
well with the predictions of the fragmentation hypothesis, whereas
in the range $5 < p_T < 10$~GeV non-fragmentation contributions
also turn out to be quite important. In effect, as shown in 
Ref.~\cite{cho}, only a linear combination of octet matrix-elements 
can be determined by fitting to the Tevatron data. However, since
the colour-octet matrix elements are universal, they will also
appear in other quarkonium production processes. So it is 
important to have independent tests of the colour-octet 
production mechanisms in other experiments \footnote{Recent papers, 
which have appeared after this talk was presented, show that a 
different linear combination of the same colour octet-matrix elements 
that appear in the Tevatron analysis also appears in the analyses 
of photoproduction \cite{photo} and hadroproduction experiments 
\cite{hadro}. These analyses provide an important cross-check on 
the colour-octet contributions.}.

Several suggestions have been made in the literature which 
could provide more stringent quantitative tests of the
colour-octet contributions and the fragmentation picture. In
particular these processes can be used to fix the colour-octet
non-perturbative matrix elements and determine the shape and size of
the fragmentation functions. The following predictions of the
theory for various experimental observables have been made:
\begin{enumerate} 
\item
Charmonium and bottomonium production at LEP provide a laboratory
for testing several features of the theory of quarkonium production.
The colour-singlet production process $Z \rightarrow J/\psi c \bar c$
\cite{lep1} turns out to be two orders of magnitude larger than the process
$Z \rightarrow J/\psi gg$ \cite{lep2} and this is due to the fact
that the former is enhanced by a fragmentation contribution which
is not suppressed by $m_c^2/m_Z^2$ \cite{brcyu}. The resulting 
prediction for prompt $J/\psi$ production is of the order of $3 \times
10^{-5}$, which is almost an order of magnitude below the experimental
number for the branching fraction obtained from LEP \cite{lep3}. 
Another fragmentation process $Z \rightarrow q \bar q g^*$ with the
virtual gluon then fragmenting into a $J/\psi$ through the colour-singlet
processes $g^* \rightarrow J/\psi gg$ and $g^* \rightarrow \chi g$ 
has been studied \cite{lep4}. Recently, the colour-octet contributions
to $J/\psi$ production in this channel have been studied \cite{lep5,lep6}
and it is found that the inclusion of the colour-octet contributions 
in the fragmentation functions results in a predictions for the 
branching ratio which is $1.4 \times 10^{-4}$ which is compatible
with the measured values of the branching fraction from LEP \cite{lep3}.
Also, the energy distribution of the $J/\psi$ coming from the colour-octet
process is very soft and is peaked at around $z=0.1$, which can be used
to distinguish it from the colour-singlet production which gives a
hard energy spectrum. The predictions from the colour-octet fragmentation 
process for the production of bottomonia \cite{lep6} are also in reasonable 
agreement with the measurements from LEP \cite{lep7}.

\item
The production of $J/\psi$ in low energy $e^+ e^-$ machines can
also provide a stringent test of the colour-octet mechanism \cite{brch}.
In this case, the colour-octet contributions dominate near the upper
endpoint of the $J/\psi$ energy spectrum, and the signature for
the colour-octet process is a dramatic change in the angular distribution
of the $J/\psi$ near the endpoint. The angular distribution has a general
form $1+A(E){\rm cos}^2 \theta$. While the colour-singlet model predicts
that $A \approx -0.84$ for $E \approx E_{\rm max}$, the colour-octet
prediction is that $A(E)$ should change sign suddenly near the endpoint
of the spectrum and approach a value of $A$ greater than $\sim +0.6$.

\item
One striking prediction of the colour-octet fragmentation process both
for $p \bar p$ colliders and for $J/\psi$ production at the $Z$-peak,
is that the $J/\psi$ coming from the process $g \rightarrow J/\psi X$
is produced in a transversely polarised state \cite{trans}. For the 
colour-octet $c \bar c$ production, this is predicted to be a 100\% 
transverse polarisation, and heavy-quark spin symmetry will then ensure 
that non-perturbative effects which convert the $c \bar c$ to a $J/\psi$
will change this polarisation only very mildly. This spin-alignment
can, therefore, be used as a test of colour-octet fragmentation.

\item
The fragmentation process has also definite predictions for the
large-$p_T$ inelastic $J/\psi$ photoproduction at HERA \cite{hera}. In the 
photoproduction process, the colour-singlet fusion process dominates
over the charm-quark and gluon fragmentation process upto a $p_T$
of 8--10~GeV. The fragmentation contributions become important
only for $p_T > 10$~GeV. Even here, it is the charm-quark fragmentation
that dominates over the gluon fragmentation process. The gluon fragmentation
process is negligibly small over the entire range of $p_T$ considered.
The $J/\psi$ cross-section at large-$p_T$ will, therefore, provide a 
handle on the charm fragmentation functions, and provide information
which is complementary to that obtained from Tevatron which is 
gluon-dominated \footnote{A recent paper \cite{cakr} deals with
photoproduction of quarkonia at HERA via colour-octet production 
processes in detail.}. 

\item
The associated production of a $J/\psi + \gamma$ also provides a
test of the fragmentation picture \cite{psigam}. This process is
dominated by the colour-singlet fusion process. If the colour-octet
fragmentation picture is the true description of the Tevatron
$J/\psi$ cross-section enhancement, then we would expect that there 
would not be a similar enhancement of the $J/\psi+\gamma$ cross-section,
and this would be very close to the colour-singlet fusion prediction
\footnote{For a more recent discussion of the colour-octet contributions to
this process, see \cite{kim}.}.

\item
Double $J/\psi$ production via double gluon fragmentation has been
studied in the context of the Tevatron \cite{bfp}. The colour-octet
fragmentation mechanism tends to increase the production rate for 
this process and the ratio of the cross-section for this process 
to that of single $J/\psi$ production is of the order of $10^{-5}$.
Given the large number of single $J/\psi$ events seen at the Tevatron,
double $J/\psi$ production may also be observable.

\item
$J/\psi$ and $\psi^{\prime}$ production in $pp$ collisions at 
centre-of-mass energies of 14~TeV at the LHC also provides a 
crucial test of colour-octet fragmentation \cite{lhc}. As in the Tevatron, 
gluon fragmentation is again the single most important source 
of quarkonium production~-- except that the magnitude of the 
gluon fragmentation contribution at the LHC completely overwhelms 
the contributions coming from either direct production or from 
charm fragmentation. Also given the large $p_T$ coverage possible 
at the LHC, the fragmentation picture becomes almost exact and 
any non-fragmentation contribution is negligibly small. An experimental 
determination of the $J/\psi$ and $\psi^{\prime}$ cross-sections at the 
LHC will provide a verification of the fragmentation picture and will 
also help to measure the colour-octet non-perturbative parameters that 
dominate the gluon fragmentation functions.
\end{enumerate} 

In summary, the theory of quarkonium production has changed
quite dramatically in an attempt to understand the anomalously
large production rates for charmonia at the Tevatron. Fragmentation
and the inclusion of colour-octet components of the quarkonium
wave-function have provided a way to resolve the discrepancy
between theory and experiment, but at the expense of introducing
undetermined non-perturbative parameters. However, these parameters
have a rigorous definition as matrix elements of colour-octet
operators in NRQCD, and are shown to be universal. This universality
helps us to predict the rates for quarkonium production processes
once these matrix-elements have been determined by fitting to one
experiment. The success of the program is, therefore, crucially dependent
on indentifying what physical observable determines a specific combination of
matrix-elements. We have reviewed some proposals in the literature,
which show how the colour-octet fragmentation picture used to 
describe the Tevatron data can be tested in other experiments. 
A preliminary study \cite{ehv} of higher-order corrections to the 
fragmentation contribution in the case of ${}^1S_0$ production has 
been made, and shows the possible importance of the higher-order 
corrections to fragmentation. More work needs to be done particularly 
in the understanding of higher-order corrections in $\alpha_s$ and $v$.

\end{document}